\documentclass[12pt]{article}
\pdfoutput=1
\usepackage{textcomp}
\usepackage{color}
\usepackage{putex}
\usepackage{autobreak}
\usepackage{indentfirst}
\usepackage{graphicx}
\usepackage{float}
\graphicspath{{plots/}}
\usepackage{tabularx}
\usepackage{caption}
\usepackage{amsmath}
\numberwithin{equation}{section}
\usepackage{cancel}
\usepackage{array}
\usepackage{subcaption}
\usepackage{epstopdf}
\usepackage{enumerate}
\usepackage{cite}
\usepackage{tensor}
\usepackage{CJKutf8}
\usepackage[utf8]{inputenc}
\usepackage{rotating}
\usepackage{multirow}
\usepackage[
      colorlinks=true,
      linkcolor=red,
      urlcolor=red,
      filecolor=black,
      citecolor=red,
      ]{hyperref}
\usepackage[noabbrev]{cleveref}
\begin{document}
%%%%%%%%%%%%%%%%%%%%%%%%%%%%%%%%%%%%%%%%%%%%%%%%%%%%%%%%%%%%%%%%%%%%%%%%%%%%%%%%%%%%%%%%%%%%%%%%%%%%%%%%%%
\newcommand{\kr}[1]{  \textbf{\textcolor{red}{(#1 --kr)}}}
\newcommand{\xz}[1]{  \textbf{\textcolor{blue}{(#1 --xz)}}}
%%%%%%%%%%%%%%%%%%%%%%%%%%%%%%%%%%%%%%%%%%%%%%%%%%%%%%%%%%%%%%%%%%%%%%%%%%%%%%%%%%%%%%%%%%%%%%%%%%%%%%%%%%
\institution{kits}{Kavli Institute for Theoretical Sciences, University of Chinese Academy of Sciences, \cr Beijing 100190, China.}
\institution{schoolphys}{School of Physical Sciences, University of Chinese Academy of Sciences, No.19A Yuquan Road, \cr Beijing 100049, China.}
%%%%%%%%%%%%%%%%%%%%%%%%%%%%%%%%%%%%%%%%%%%%%%%%%%%%%%%%%%%%%%%%%%%%%%%%%%%%%%%%%%%%%%%%%%%%%%%%%%%%%%%%%%
\title{Spin-2 operators in two-dimensional $\mathcal{N}=(4,0)$ quivers from massive type IIA}
%%%%%%%%%%%%%%%%%%%%%%%%%%%%%%%%%%%%%%%%%%%%%%%%%%%%%%%%%%%%%%%%%%%%%%%%%%%%%%%%%%%%%%%%%%%%%%%%%%%%%%%%%%
\authors{
Shuo Zhang\worksat{\kits}${}^{,}$\worksat{\schoolphys}${}^{,}$\footnote{{\hypersetup{urlcolor=black}\href{mailto:zhangshuo163@mails.ucas.ac.cn
}{zhangshuo163@mails.ucas.ac.cn}}}
}
%%%%%%%%%%%%%%%%%%%%%%%%%%%%%%%%%%%%%%%%%%%%%%%%%%%%%%%%%%%%%%%%%%%%%%%%%%%%%%%%%%%%%%%%%%%%%%%%%%%%%%%%%%
\abstract{
In this work we revisit the problem of studying spin-2 fluctuations around a class of solutions in massive type IIA that is given by a warped $\text{AdS}_3 \times \text{S}^2 \times \text{T}^4 \times \mathcal{I}_{\rho}$ and with $\mathcal{N}=(4,0)$ supersymmetry. We were able to identify a class of fluctuations, which is known as the ``minimal universal class'' that is independent of the background data and saturates the bound on the mass related to the field theory unitarity bound. These operators have conformal dimension $\Delta = 2(\ell+1)$, with $\ell$ being the quantum number of the angular momentum on the $\text{S}^2$. We also computed the normalisation of the $2$-point function of stress-energy tensors from the effective $3$-dimensional graviton action. We comment on the relation of our results to the related $\text{AdS}_3$ and $\text{AdS}_2$ solutions in massive type IIA and type IIB theories respectively. 
}
\date{\today}
%%%%%%%%%%%%%%%%%%%%%%%%%%%%%%%%%%%%%%%%%%%%%%%%%%%%%%%%%%%%%%%%%%%%%%%%%%%%%%%%%%%%%%%%%%%%%%%%%%%%%%%%%%
\maketitle
%%%%%%%%%%%%%%%%%%%%%%%%%%%%%%%%%%%%%%%%%%%%%%%%%%%%%%%%%%%%%%%%%%%%%%%%%%%%%%%%%%%%%%%%%%%%%%%%%%%%%%%%%%
{
\hypersetup{linkcolor=black}
\tableofcontents
}
%%%%%%%%%%%%%%%%%%%%%%%%%%%%%%%%%%%%%%%%%%%%%%%%%%%%%%%%%%%%%%%%%%%%%%%%%%%%%%%%%%%%%%%%%%%%%%%%%%%%%%%%%%
\newpage
%%%%%%%%%%%%%%%%%%%%%%%%%%%%%%%%%%%%%%%%%%%%%%%%%%%%%%%%%%%%%%%%%%%%%%%%%%%%%%%%%%%%%%%%%%%%%%%%%%%%%%%%%%
\section{Introduction}\label{sec: intro}
Two-dimensional conformal field theories (CFTs), with and without supersymmetry, have been studied extensively, since they play a prominent role in various different physics contexts ranging from string theory to condensed matter systems and black holes. They have two main characteristics that distinguish them from their higher-dimensional cousins. Firstly, the conformal group in two dimensions is infinite dimensional\cite{Belavin:1984vu}. Furthermore, in certain examples in two dimensions, the theories are exactly solvable owing to the presence of integrability \cite{DiFrancesco:1997nk}. 

Due to the holographic principle\cite{Maldacena:1997re}, the study of two-dimensional CFTs can be equivalently phrased as the study of various aspects of supergravity theories defined in an $\text{AdS}_3$ spacetime. 

It follows from the above that studying two-dimensional superconformal field theories and their dual AdS gravitational description provides us with a theoretical lab to test various ideas and aspects of the AdS/CFT duality itself. As a matter of fact, specific $\text{AdS}_3$  backgrounds have fields only in the NS-NS sector, and provide us with exactly solvable backgrounds in string theory\cite{Maldacena:2000hw,Maldacena:2000kv}.

The first example of an $\text{AdS}_3$ solution, in the above context, is the background that appears as the near-horizon limit of D$1$- and D$5$-branes. This gives rise to an $\text{AdS}_3 \times \text{S}^3 \times \text{CY}_2$ background\cite{Maldacena:1997re}. For the case of $\text{CY}_2 = \text{K}3$ or $\text{T}^4$ much progress has been made both from calculations in the the supergravity \cite{Deger:1998nm,deBoer:1998us,Giveon:1998ns,deBoer:1998kjm,Hampton:2018ygz,Gaberdiel:2018rqv,Eberhardt:2018ouy,Eberhardt:2019qcl,Keller:2019suk} and the field theory sides \cite{Bombini:2017sge,Giusto:2018ovt,Bombini:2019vnc,Roumpedakis:2018tdb,Rastelli:2019gtj,Aprile:2021mvq,Galliani:2017jlg,Giusto:2019pxc}.

Despite the tremendous progress reported in the above works related to the study of the D$1$/D$5$ background and its dual CFT, until recently not a lot was known with regard to the holographic realization of two-dimensional conformal field theories with less supersymmetry. This motivated the authors of \cite{Lozano:2019emq} to look for and obtain a complete classification of $\text{AdS}_3$ solutions in massive type IIA that preserves $\mathcal{N}=(4,0)$ supersymmetry\footnote{A related study to the classification of $\text{AdS}_3$ backgrounds is the recent result of \cite{Legramandi:2020txf} that offers a complete classification of $\text{AdS}_3$ solutions in $10$ and $11$ dimensions with $\mathcal{N}=(8,0)$ supersymmetry.}. Subsequently \cite{Lozano:2019jza,Lozano:2019zvg,Lozano:2019ywa,Filippas:2019ihy,Speziali:2019uzn,VanGorsel:2019xcw,Filippas:2020qku,Zacarias:2021pfz}, we learned a great deal about this setup, as well as other $\text{AdS}_3$ backgrounds with less than $\mathcal{N}=(4,4)$ supersymmetry \cite{Couzens:2019iog,Couzens:2019mkh,Legramandi:2019xqd,Faedo:2020nol,Faedo:2020lyw,Couzens:2021tnv,Macpherson:2021lbr,Macpherson:2022sbs,Couzens:2022agr,Conti:2023rul,Passias:2019rga,Passias:2020ubv}.

Related to a better understanding of the theories developed in \cite{Lozano:2019emq} is the spectrum of operators in their dual field theory descriptions. By using the holographic dictionary we know the spectrum describing the gauge invariant field theory operators is derived upon considering the linearized fluctuations in the bulk supergravity. While this is conceptually straightforward, performing the computation and subsequent analysis of the complete set of linearized fluctuations for a given supergravity background is a daunting task. This is why, until this day we only have a handful of results, see e.g. \cite{PhysRevD.32.389,Deger:1998nm,Eberhardt:2017fsi}.

Despite the objective difficulty in this line of analysis, the work of \cite{Bachas:2011xa} concluded that if we focus only on the spin-2 operators and their spectrum the computation simplifies significantly. Particularly, the authors in \cite{Bachas:2011xa} proved that fluctuations along the non-compact part of the metric satisfy an equation defined purely in terms of the geometrical data of the background supergravity solution. These metric fluctuations are the ones that are dual to spin-2 field theory states.

The authors of the above paper performed their analysis for an $\text{AdS}_4$ solution, however, a generalization of their logic to arbitrary space-time dimensions is straightforward. This paved the way for many new and exciting results related to the spectrum of holographic spin-2 operators across different dimensions \cite{Rigatos:2022ktp,Speziali:2019uzn,Lima:2022hji,Klebanov:2009kp,Itsios:2019yzp,Chen:2019ydk,Gutperle:2018wuk,Passias:2018swc,Passias:2016fkm,Roychowdhury:2023lxk,Apruzzi:2019ecr,Apruzzi:2021nle,Lima:2023ggy,Zhang:2024pim}. 

In these works the authors have been able to identify a particular class of supergravity mode solutions, the so-called ``universal minimal class''. The name is justified by the fact that it does not depend on the specific details of a given supergravity background, and is dubbed minimal because it saturates the unitarity bound. 

In this work we are going to continue this line of analysis by examining a warped class of background solutions of the form $\text{AdS}_3 \times \text{S}^2 \times \text{T}^4 \times \mathcal{I}_{\rho}$ that was derived in \cite{Lozano:2019jza}. While this family of solutions was considered from the point of view of the spin-2 analysis in \cite{Speziali:2019uzn}, only one of the two choices of the solutions to the BPS condition was studied. In this work we consider the second possibility. We will clarify and make these two choices explicit in the next section. 

\begin{figure}[H]
\centering
\includegraphics[width=0.75\textwidth]{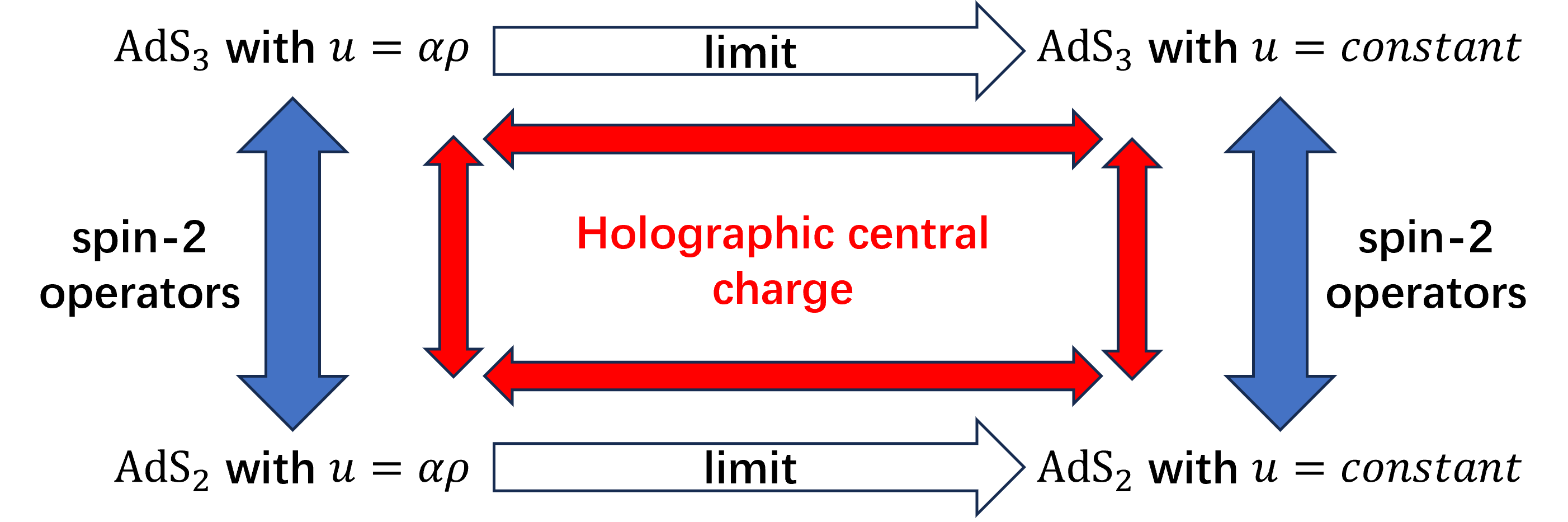}
\caption{A pictorial depiction of the map among the different choices of the solution of the BPS condition for the $\text{AdS}_3$ backgrounds developed in \cite{Lozano:2019jza} and their related $\text{AdS}_2$ backgrounds developed in \cite{Lozano:2020txg} obtained by a T-duality. A double arrow means that the corresponding quantities are the same in those setups, while a single arrow means that an appropriate limit has to be considered.}
\label{fig：holographic central charge}
\end{figure}

One motivation of pursuing our line of analysis, is that these two choices for the solution of the BPS condition lead to different setups, so it is reasonable to ask the question of the spectrum, spin-2 supergravity modes and central charge in both cases. Another motivation is coming from recent results in a related class, via a T-duality transformation, of $\text{AdS}_2$ backgrounds \cite{Rigatos:2022ktp,Zhang:2024pim}. In those works, when pursuing the spin-2 analysis there were some similarities and some differences. This naturally leads to our final motivation, which is to establish a map among the different choices for the solution of the BPS condition and the related T-dual backgrounds that we schematically present in \cref{fig：holographic central charge}

The outline of this paper is the following: \cref{sec: bg} is a review of the class of supergravity backgrounds that were derived in \cite{Lozano:2019jza} and their dual field theory realizations. In  \cref{sec: Fl} we study the linearized fluctuations along the AdS directions and derive the equation of motion they satisfy. In \cref{sec: ub&us} we calculate the unitarity bound and the so-called ``minimal universal class of solutions''. \Cref{sec: cc} contains our computation of the holographic central charge and in \cref{sec: multiplets} we discuss some field theory aspects of our computations. We conclude and offer some thoughts for future research topics in \cref{sec: final}.

\section{Preliminaries}\label{sec: bg}

The purpose of this section is to review the main aspects of the global class of supergravity solutions that was constructed in \cite{Lozano:2019jza}. Additionally, we discuss the main characteristic of their dual field theory realisations.

\subsection{The supergravity background}

The NS-NS sector of the solution of \cite{Lozano:2019jza} consists of the following metric:
\begin{equation}\label{eq: nsns_sector_metric}
ds^2 = \frac{u}{\sqrt{h_4 h_8}} \left(ds^2_{\text{AdS}_3} + \frac{1}{4} ds^2_{\text{S}^2} \right) + \sqrt{\frac{h_4}{h_8}}ds^2_{\text{T}^4} + \frac{\sqrt{h_4 h_8}}{u} d\rho^2 
\,          ,
\end{equation}
where the metric of the torus is given by:
    \begin{equation}
        \text{d}s^2_{\text{T}^4} = \text{d}\theta_1^2+\text{d}\theta_2^2+\text{d}\theta_3^2+\text{d}\theta_4^2,
    \end{equation}
This sector is, also, supplemented by the following dilaton and the three-form flux:
\begin{equation}\label{eq: nsns_sector_full}
    \begin{aligned}
e^{-2\phi} &= h_8^{5/2} \sqrt{\frac{h_4}{u}}
\,              ,
\\
H &= \frac{1}{2} \vol_{\text{S}^2} \wedge d\rho
\,              .
    \end{aligned}
\end{equation}

The R-R sector reads \cite{Lozano:2019jza}:
\begin{equation}\label{eq: rr_sector_full}
    \begin{aligned}
    F_0 &= h^{\prime}_8
    \,          ,
    \\
    F_2 &= - \frac{1}{2} h_8 \vol_{\text{S}^2}
    \,          ,
    \\
    F_4 &= 2 h_8 d\rho \wedge \vol_{\text{AdS}_3} - \partial_{\rho}h_4 \vol_{\text{T}^4}
    \,          .
    \end{aligned}
\end{equation}
The remaining R-R fluxes can be obtained in a straightforward manner from the above expressions by considering their ten-dimensional Hodge duals. In particular we have $\star_{10} F_0 = - F_{10}, ~ \star_{10} F_2 =  F_{8}, ~ \star_{10} F_4 = - F_{6}$. In the above equations in principle there can be a term proportional to $H_2$. This $H_2$ is given in terms of three arbitrary functions defined on the $\text{T}^4$ and is a 2-form in terms of the one-form fields in the canonical frame of the torus; for its explicit form see \cite[eq.(3.5)]{Lozano:2019emq}. This would add a term in the NS-NS three-form flux and an additional term in the two-form flux in the R-R sector. Whether this term is present or not does not affect our computations and we chose to disregard it, in order to simplify our presentation.

The BPS equations lead to the condition\footnote{In the case that $H_2$ is non-vanishing we have a second condition coming from supersymmetry which is: $H_2 + \star_{4} H_2 = 0$, with $\star_{4}$ being the Hodge dual defined on the torus.}
\begin{equation}\label{eq: bps_eq}
    u^{\prime \prime}(\rho) = 0
    \,      ,
\end{equation}
while the Bianchi identities yield:
\begin{equation}\label{eq: aux_rev_01}
h^{\prime \prime}_4(\rho) = 0 \,      ,       \qquad      h^{\prime \prime}_8(\rho) = 0
\,          .
\end{equation}

It is immediately clear from \cref{eq: bps_eq} that there are two choices for the characteristic function $u(\rho)$ presented below: 
\begin{equation}
u = \frac{u_0}{2\pi} \rho
\,              ,
\qquad
u = u_0
\,              ,
\qquad
\text{with} \qquad u_0 = \texttt{constant}
\,          .
\end{equation}  

In our presentation of the background, see \cref{eq: nsns_sector_metric,eq: nsns_sector_full,eq: rr_sector_full}, we have already imposed that $u = u_0$ which is the case that we study in this work. This sets to zero some derivative terms in those expressions that are otherwise present.

It is worth noting that \cref{eq: aux_rev_01} holds true in the absence of explicit brane sources. However, violations of the Bianchi identities can occur at specific points where explicit brane sources are present. This results in a modification of the right-hand side of \cref{eq: aux_rev_01} through the inclusion of contributions from $\delta$-functions, leading to the characterization of the functions $h_4$ and $h_8$ as piecewise linear functions.

Our focus in this study is on an infinite class of solutions defined in such a way that these characteristic functions are piecewise continuous and vanish at the endpoints of the $\mathcal{I}_{\rho}$.  Explicitly they are given by \cite{Lozano:2019jza}: 

\iffalse
    \begin{equation}\label{eq: nsns_sector_full}
    \begin{aligned}
         \text{d}s^2 &=\frac{u}{\sqrt{h_4h_8}}\left(\text{d}s^2_{\text{AdS}_3}+\frac{h_4h_8}{4h_4h_8+(u')^2}\text{d}s^2_{\text{S}^2}\right)+\sqrt{\frac{h_4}{h_8}}\text{d}s^2_{\text{T}^4}+\frac{\sqrt{h_4h_8}}{u}\text{d}\rho^2,\\
         e^{-2\Phi} &=\frac{h_8}{4u}\sqrt{\frac{h_8}{h_4}}(4h_4h_8+(u')^2),\\
         H &=\frac{1}{2}\text{vol}_{\text{S}^2}\wedge\text{d}\left(\rho-\frac{uu'}{4h_4h_8+(u')^2}\right)+\frac{1}{h_8}\text{d}\rho\wedge H_2,
    \end{aligned}
    \end{equation}

    \begin{equation}
     h''_4(\rho)=0, \qquad h''_8(\rho)=0
     \,             ,
    \end{equation}
    and
    \begin{equation}
        u''(\rho)=0.
    \end{equation}
        \begin{equation}\label{eq: metrics}
    \begin{aligned}
        &\text{d}s^2=\frac{u}{\sqrt{h_4h_8}}(\text{d}s^2_{\text{AdS}_3}+\frac{1}{4}\text{d}s^2_{\text{S}^2})+\sqrt{\frac{h_4}{h_8}}\text{d}s^2_{\text{T}^4}+\frac{\sqrt{h_4h_8}}{u}\text{d}\rho^2,\\
        &e^{-2\Phi}=h_8^2\sqrt{\frac{h_4 h_8}{u}},\\
        &H=\frac{1}{2}\text{vol}_{\text{S}^2}\wedge\text{d}\rho+\frac{1}{h_8}\text{d}\rho\wedge H_2,
    \end{aligned}
    \end{equation}
    \fi 

    \begin{equation}
    h_4(\rho)=\Upsilon
    \begin{cases}
    \beta_0\frac{\rho}{2\pi} &\quad 0\leq\rho\leq2\pi\\
    \sum\limits_{i=0}\limits^{k-1}\beta_i+\beta_k\left(\frac{\rho}{2\pi}-k\right) &\quad 2\pi k<\rho\leq2\pi(k+1),\quad k=1,\cdots,P-1\\
    \alpha_P+\alpha_P\left(-\frac{\rho}{2\pi}+P\right) &\quad 2\pi P<\rho\leq2\pi(P+1),
    \end{cases}
    \end{equation}
        \begin{equation}
    h_8(\rho)=
    \begin{cases}
    \nu_0\frac{\rho}{2\pi} &\quad 0\leq\rho\leq2\pi\\
    \sum\limits_{i=0}\limits^{k-1}\nu_i+\nu_k\left(\frac{\rho}{2\pi}-k\right) &\quad 2\pi k<\rho\leq2\pi(k+1),\quad k=1,\cdots,P-1\\
    \mu_P+\mu_P\left(-\frac{\rho}{2\pi}+P\right) &\quad 2\pi P<\rho\leq2\pi(P+1).
    \end{cases}
    \end{equation}
In the relations presented above, $\Upsilon$ represents an overall normalization 
%(see e.g. \cite{Lozano:2019jza} for the exact relation $\hat{h}_4 = \Upsilon h_4$). 
The integration constants $\alpha_P,\beta_k,\mu_P,\nu_k$ with $(k=0,\cdots,P-1)$ are of interest. Demanding that the functions are continuous functions along the interval defined by the $\rho$-coordinate, we can conclude that \cite{Lozano:2019jza}:    
    \begin{equation}
    \alpha_P=\sum_{i=0}^{k-1}\beta_i,\qquad \mu_P=\sum_{i=0}^{k-1}\nu_i.
    \end{equation}
    
We are specifically investigating cases where the $\rho$-coordinate defines a finite interval denoted by $\mathcal{I}_{\rho}$ within the range $[0,\rho^\ast]$, where $\rho^\ast$ is conveniently set to $\rho^\ast=2\pi(P+1)$ for a large integer $P$ \cite{Lozano:2019jza}. In the class of solutions relevant to our studies, it is essential that the functions $h_4(\rho)$ and $h_8(\rho)$ vanish at the endpoints of the interval $\mathcal{I}_{\rho}$, as we have already mentioned, and specifically we have $h_4(0)=h_8(0)=h_4(\rho^\ast)=h_4(\rho^\ast)=0$\footnote{We describe other viable possibilities in \cref{sec: final}. These have been carefully scrutinized in \cite{Filippas:2020qku}.}. 
%We illustrate this behavior in \cref{fig: h4h8uplot}:
%    \begin{figure}[H]
%\centering
%\includegraphics[width=0.75\textwidth]{shuo}
%\caption{An example of the characteristic functions $h_4(\rho), h_8(\rho)$, as well as the function $u$, where the functions $h_4(\rho)$ and $h_8(\rho)$ vanish at the edges of the interval $\mathcal{I}_{\rho}$ and $u$ is a constant.}
%\label{fig: h4h8uplot}
%\end{figure}

The global class of supergravity backgrounds that we described above can be pictorially represented by the brane realization in \cref{branes living space}.
        \begin{table}[H]
	\begin{center}
		\begin{tabular}{|c|c|c|c|c|c|c|c|c|c|c|c|}
			\hline	
 &&&&&&&&&&\\[-0.95em] 	    
	    &	$x^0$	& $x^1$ 		& $x^2$ 		& $x^3$			& $x^4$ 		& $x^5$ 		& $x^6$ 		& 	$x^7$ 		& 	$x^8$ 		& 	$x^9$ 			\\ \hline \hline
D2 		& 	--- 	& ---		    & $\bullet$		& $\bullet$ 	& $\bullet$ 	& $\bullet$ 	& ---  	        &   $\bullet$	& 	$\bullet$  	&  $\bullet$ 		\\ \hline
D4 	    & 	---		& ---   		& $\bullet$ 	& $\bullet$ 	& $\bullet$ 	& $\bullet$		& $\bullet$ 	& 	--- 		& 	--- 		&  --- 		\\ \hline
D6 		& 	---		& --- 			& 	--- 		& --- 			& --- 			& --- 			& ---         	&   $\bullet$	&   $\bullet$	&  $\bullet$ 		\\ \hline
D8 		& 	---		& ---			&	---  		& --- 			& --- 			& ---		    & $\bullet$  	& 	--- 		& 	--- 		&  --- 		\\ \hline
NS5 	& 	---		& ---			&	---  		& --- 			& --- 			& ---		    & $\bullet$  	&   $\bullet$	&   $\bullet$	& 	$\bullet$  			\\ \hline
		\end{tabular} 
\caption{The $\tfrac{1}{4}$-BPS brane set-up of the supergravity backgrounds. We have used (---) to denote that the brane lies in that direction. A ($\bullet$) denotes a transverse coordinate to the brane. The dimensions $x^0,x^1$ represent the two-dimensional Minkowski spacetime where the field theory lives. The $x^2,x^3,\ldots,x^5$ directions are those of the $\text{T}^4$, where the D$6$ and D$8$ are wrapped. The $x^6$ coordinate is the $\rho$-dimension. The remaining three dimensions realize the $\text{SO}(3)$ symmetry associated with the $\text{S}^2$.}
\label{branes living space}
\end{center}
\end{table}

In \cref{branes living space}, the dimensions $x^0$ and $x^1$ correspond to the two-dimensional Minkowski spacetime where the field theory is defined. The directions $x^2, x^3, x^4, x^5$ represent the $\text{T}^4$, around which the D$6$ and D$8$ are wrapped. The coordinate $x^6$ corresponds to the $\rho$-dimension. The remaining three dimensions realize the $\text{SO}(3)$ symmetry associated with the $\text{S}^2$.

\subsection{The field theory picture}
\begin{figure}[H]
\centering
\includegraphics[width=0.75\textwidth]{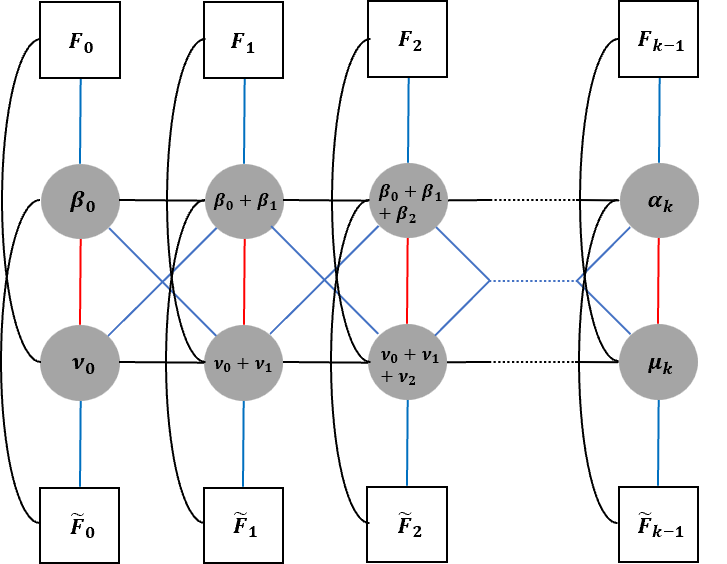}
\caption{The general 2d quiver field theory that is the seed of the CFT$_1$ considered in this work. The black line denotes a (4,4) hypermultiplet, the red line represents a (4,0) hypermultiplet and the blue line a (2,0) multiplet. Each gauge node has $\mathcal{N}=(4,4)$ vector multiplets degree of freedom.}
\label{fig: quiver plot}
\end{figure}

This section will cover the fundamental aspects of the CFT that is dual to the AdS$_3$ background discussed in \cref{eq: nsns_sector_metric,eq: nsns_sector_full,eq: rr_sector_full}. It was argued in \cite{Lozano:2019zvg} that these AdS$_3$ solutions are dual to the IR fixed-point limit of the quiver theory presented in \cref{fig: quiver plot}.

More precisely, in \cite{Lozano:2019zvg} it was shown that this class of warped AdS$_3$ solutions are the dual descriptions of the IR fixed-point limit of the quiver theory presented in \cref{fig: quiver plot}. Particularly, the quiver diagram presented in \cref{fig: quiver plot} flows to a fixed-point solution in the IR whose dynamics is captured by the AdS$_3$ background solution of \cite{Lozano:2019jza}. We note that the work of \cite{Couzens:2021veb} provides us with the most up-to-date analysis on the aspects of the two-dimensional CFT. More importantly it corrects some details that were overlooked in the preliminary field analysis of \cite{Lozano:2019jza,Lozano:2019zvg}. In our presentation of the matter content below, we closely follow the full-fledged analysis of \cite{Couzens:2021veb}. 

The basic building blocks of the field theory as they are depicted in the quiver diagram, see \cref{fig: quiver plot}, are the following: a circle denotes a gauge node in the quiver with the associated group being the special unitary group. Therefore, the $\beta_0$ in the circle means $\text{SU}(\beta_0)$ and similarly for the other nodes of this kind. Additionally, we have used squares to stand for the flavor groups, and hence a square with $F_0$ is the flavor group $\text{SU}(F_0)$. The black lines connecting the various gauge nodes are $\mathcal{N}=(4,4)$ hypermultiplets and the red lines represent $\mathcal{N}=(4,0)$ hypermultiplets. We have used blue lines to denote $\mathcal{N}=(2,0)$ Fermi multiplets between two gauge nodes. Each gauge node comes equipped with $\mathcal{N}=(4,4)$ vector multiplets \footnote{For details on the basic aspects of $2$-dimensional $\mathcal{N}=(4,0)$ and $\mathcal{N}=(2,0)$ multiplets we refer the interested reader to either \cite[appendix B]{Speziali:2019uzn} or \cite[appendix B]{Filippas:2020qku} and references therein.}. 

It is important to note the two rows of flavour groups, and also the two rows of the color nodes. This is crucial because it implies that the $F$'s and $\tilde{F}$'s are not independent of the number of colors in the quiver. This is required for the consistency of the field theory. The reason for this is that the field theory is a chiral theory. Therefore, we have to ensure that gauge anomalies are cancelled. This has been explicitly studied in \cite{Lozano:2019zvg} and the cancellation of gauge anomalies leads to the precise condition:
\begin{equation}\label{eq: anomalycancellation}
    F_0 = \nu_0 - \nu_1
    \,      ,
    \qquad
    \tilde{F}_0 = \beta_0 - \beta_1
    \,      .
\end{equation}

The relation presented above, \cref{eq: anomalycancellation}, is an application of cancelling the gauge anomalies at the first node of the quiver only. However, the same treatment has to be carried out at each node of the quiver and this leads to similar relations among the characteristic numbers of the groups. Carrying out the full computation was done explicitly in \cite{Lozano:2019zvg} leading to:
\begin{equation}\label{eq: anomalycancellationfinal}
    F_{k-1} = \nu_{k-1} - \nu_k
    \,      ,
    \qquad
    \tilde{F}_{k-1} = \beta_{k-1} - \beta_k
    \,      .
\end{equation}

A final comment before closing off this section and proceeding with our analysis of the metric fluctuations is that the numbers appearing in \cref{eq: anomalycancellationfinal} are the number of color and flavor branes in the bulk supergravity description. 

\section{AdS metric fluctuations}\label{sec: Fl}
It is more convenient for our purposes to bring the metric to the Einstein frame. To do so, we multiply the line element by $e^{-\frac{\Phi}{2}}$, where $\Phi$ is the dilaton field. Hence, we have: 
\begin{equation}
        \text{d}s^2=e^{-\frac{\Phi}{2}}f_1\text{d}s^2_{\text{AdS}_3}+\hat{g}_{ab}\text{d}z^a\text{d}z^b
        \,      ,
\end{equation}
with the internal metric given by:
\begin{equation}
        \hat{g}_{ab}\text{d}z^a\text{d}z^b=e^{-\frac{\Phi}{2}}\left(f_1\text{d}s^2_{\text{S}^2}+\sqrt{\frac{h_4}{h_8}}\text{d}s^2_{\text{T}^4}+\frac{\sqrt{h_4h_8}}{u}\text{d}\rho^2\right)
        \,      ,
\end{equation}
and we have used the shorthand $f_1$ to denote
    \begin{equation}
        f_1 = \frac{u}{\sqrt{h_4h_8}}
        \,   .
    \end{equation}

%Our approach is to begin with the perturbations of the metric. For the sake of our analysis, we find it most convenient to rewrite the metric in the Einstein frame. To achieve this, we multiply it by a factor of the dilaton, more precisely,
%$e^{-\frac{\Phi}{2}}$. 

%    \begin{equation}
%    \begin{aligned}
%        \hat{g}_{ab}\text{d}z^a\text{d}z^b&=e^{-\frac{\Phi}{2}}\left(\frac{u}{4\sqrt{h_4h_8}}\text{d}s^2_{\text{S}^2}+\sqrt{\frac{h_4}{h_8}}\text{d}s^2_{\text{T}^4}+\frac{\sqrt{h_4h_8}}{u}\text{d}\rho^2\right)
%        \,      ,
%        \\
%        f_1&=\frac{u}{\sqrt{h_4h_8}}\,   ,
%    \end{aligned}
%    \end{equation}
%    such that the metric in the Einstein frame becomes:
%    \begin{equation}
%        \text{d}s^2=e^{-\frac{\Phi}{2}}f_1\text{d}s^2_{\text{AdS}_3}+\hat{g}_{ab}\text{d}z^a\text{d}z^b.
%    \end{equation}
We are focusing on the fluctuations of the metric, which we denote as by $h$, along the $\text{AdS}_3$ directions of the 10-dimensional background. Specifically:
    \begin{equation}
        \text{d}s^2=e^{-\frac{\Phi}{2}}f_1(\text{d}s^2_{\text{AdS}_3}+h_{\mu\nu}(x,z)\text{d}x^\mu\text{d}x^\nu)+\hat{g}_{ab}\text{d}z^a\text{d}z^b.
    \end{equation}
    
We separate the variables in the mode describing the fluctuations and decompose $h_{\mu\nu}(x,z)$ into two bits: one describing the transverse-traceless component in the bulk and another one defined on the internal manifold and represented by a  scalar function, namely 
    \begin{equation}
        h_{\mu\nu}(x,z)=h_{\mu\nu}^{(tt)}(x)\psi(z).
    \end{equation}
The transverse-traceless tensor $h_{\mu\nu}^{(tt)}(x)$ satisfies:
    \begin{equation}\label{eq: spin_2_ads_3_eigen}
        \Box_{\text{AdS}_3}^{(2)}h_{\mu\nu}^{(tt)}(x)=(M^2-2)h_{\mu\nu}^{(tt)}(x)=(\Box_{\text{AdS}_3}-2)h_{\mu\nu}^{(tt)}(x).
    \end{equation}
The $\Box_{\text{AdS}_3}^{(2)}$ here is the $\text{AdS}_3$ Laplacian acting on a rank-2 tensor field, and we have used a well-known identity to relate it to $\Box_{\text{AdS}_3}$, which is the oridnary scalar Laplacian in $\text{AdS}_3$ \cite{Polishchuk:1999nh}.

We are now at a position to use one of the main results of  \cite{Bachas:2011xa}. In that work, the authors showed that the linearized Einstein equations can be simplified to a ten-dimensional massless Klein-Gordon problem for $h_{\mu \nu}$. In particular, our starting point is:    
    \begin{equation}\label{eq: laplacian_01}
        \frac{1}{\sqrt{-g}}\partial_M\sqrt{-g}g^{MN}\partial_Nh_{\mu\nu}=0.
    \end{equation}
We note that the capital indices ${M,N}$ represent directions in the ten-dimensional background. They naturally split into $M=\{\mu,a\}$, with $\mu$ representing an index in the non-compact spacetime and the index $a$ representing an index along the compact directions. After writing out \cref{eq: laplacian_01} explicitly, we obtain:
    \begin{equation}    
        \hat{\Box}\psi(z)=-M^2\psi(z)        
        \,      ,
    \end{equation}
    where we have defined the operator $\hat{\Box}$ to be given by:  
    \begin{equation}\label{eq: hatbox_def}
        \hat{\Box}\equiv\frac{1}{\sqrt{\hat{g}}}\partial_a(f_1e^{-\frac{\Phi}{2}}\sqrt{\hat{g}}\hat{g}^{ab}\partial_b)
        \,          ,
    \end{equation}
    and we have used \cref{eq: spin_2_ads_3_eigen} in the above derivation. This means that after using the eigenvalues for the mode in the 3-dimensional AdS subspace, we have simplified the 10-dimensional equations of motion to a 7-dimensional dynamical problem with a mass term. We can evaluate the contributions from the different parts of the internal space and arrive at:
    \begin{equation}\label{eq: eom_aux_1}
        \left(4\nabla_{\text{S}^2}+\frac{u}{h_4}\sum_{i=1}^n\partial_{\theta_i}^2+\frac{u^2}{h_4h_8}\partial_\rho^2+M^2\right)\psi=0.
    \end{equation}
    
In order to solve the differential equation given by \cref{eq: eom_aux_1}, it is most useful to decompose the internal component of the metric fluctuation, denoted as $\psi$, into its respective eigenstates pertaining to the distinct segments of the internal space. This decomposition reads:   \begin{equation}\label{eq: mode_expansion}
        \psi=\sum_{\ell mn}\Psi_{\ell mn}Y_{\ell m}e^{in\cdot\theta}
        \,      ,
    \end{equation}
where in the above $n\cdot\theta=n_1\theta_1+n_2\theta_2+n_3\theta_3+n_4\theta_4$. Upon inserting the decomposition provided by \cref{eq: mode_expansion} into \cref{eq: eom_aux_1}, we derive the following:
    \begin{equation}\label{eq: ODE}
        \frac{\text{d}^2}{\text{d}\rho^2}\Psi-\frac{h_8n^2}{u}\Psi+\frac{M^2-4\ell(\ell+1)}{u^2}h_4h_8\Psi=0
        \,      ,
    \end{equation}
where in the above expression, we have omitted the quantum numbers as subscripts for the sake of simplicity; thus, $\Psi_{\ell mn}$ is represented as $\Psi$. 

This ordinary differential equation is a Sturm-Liouville problem. We can write it in a more suggestive way, namely in the standard way of expressing any Sturm-Liouville problem. This is done in the following way:    
    \begin{equation}\label{sturm_liouville standard form}
        \frac{\text{d}}{\text{d}\rho}\left(L(\rho)\frac{\text{d}}{\text{d}\rho}\right)\Psi+Q(\rho)\Psi=-\lambda W(\rho)\Psi
        \,          ,
    \end{equation}
    with the characteristic quantities being given by:     
        \begin{equation}\label{eq: sturm_liouville}
        \begin{aligned}
        L(\rho)&=1\,    , 
        \\
        Q(\rho)&=\frac{h_8n^2}{u}\,   , 
        \\
        W(\rho)&=h_4h_8\,   , 
        \\
        \lambda&=\frac{M^2-4\ell(\ell+1)}{u^2}\,      .
    \end{aligned}
    \end{equation}
%The eigenvalues denoted by $\lambda$ are integral to the Sturm-Liouville problem. The parameter $\rho$ ranges from 0 to $2\pi(P+1)$, with $P$ representing a large integer. It is important to note that the defining functions vanish at the boundaries of $\mathcal{I}_{\rho}$, a characteristic defining a Sturm-Liouville problem in mathematical literature. Our subsequent section will delve into the requisite boundary conditions essential for the comprehensive definition of the Sturm-Liouville problem, along with the derivation of its solutions.    
    
\section{Unitarity boundary and universal solutions}\label{sec: ub&us}
The next step in our analysis is to obtain the unitarity bound for these operators. To do so, we first need to obtain an inequality for the mass eigenvalue in terms of the angular momentum, $\ell$. Then, by utilizing the standard AdS/CFT formula, $M^2 = \Delta(\Delta-2)$, we can relate the conformal dimension of the field theory operators to the value of $\ell$. 

We begin the derivation of the mass inequality by multiplying the Sturm-Liouville problem, see \cref{eq: ODE}, by $\Psi$. We obtain: 

%In the scope of our current work, our objective is to establish a constraint for the mass eigenvalue $M^2$, originating from the Sturm-Liouville problem presented in the preceding section (refer to equation \cref{eq: ODE}). Subsequently, leveraging the foundational holographic relationship between mass and conformal dimension, we aim to deduce the unitarity constraint. Our approach involves acquiring an inequality for $M^2$ from the equation of motion, focusing specifically on a category of solutions documented in existing literature as the ``minimal universal class of solution''. This class of solutions exhibits universality by being independent of the precise forms of $h_4$ and $h_8$.

%Commencing this derivation of the mass inequality, which will ultimately lead to the unitarity constraint, entails the multiplication of the Sturm-Liouville problem, as defined in \cref{eq: ODE}, by $\Psi$, followed by integration across domain $\mathcal{I}_{\rho}$. This process yields:

\begin{equation}
        \int_{\mathcal{I}_\rho}\Psi\frac{\text{d}^2}{\text{d}\rho^2}\Psi \text{d}\rho-\int_{\mathcal{I}_\rho}\frac{h_8n^2}{u}\Psi^2\text{d}\rho+\frac{M^2-4\ell(\ell+1)}{u^2}\int_{\mathcal{I}_\rho}h_4h_8\Psi^2\text{d}\rho=0.
\end{equation}
In the given interval $\mathcal{I}_{\rho}$, it is crucial to note that the upper bound, denoted as $\rho^\ast$, is greater than zero. We focus on the term with the second derivative and perform an integration by parts. This yields:

%Our initial focus lies on the term containing the second derivative of $\Psi$. Employing integration by parts facilitates the acquisition of the following expression:
    
\begin{equation}\label{integration equation}
        \frac{M^2-4\ell(\ell+1)}{u^2}\int_0^{\rho^\ast}h_4h_8\Psi^2\text{d}\rho+\Psi\Psi'\big|_0^{\rho^\ast}=\int_0^{\rho^\ast}\left((\Psi')^2+\frac{h_8n^2}{u}\Psi^2\right)\text{d}\rho.
\end{equation}
Note that in the above we have denoted by $\Psi'$ the derivative of $\Psi$ with respect to $\rho$. 

%preceding discussion, it is observed that $\Psi'$ represents the derivative of $\Psi$ with the variable $\rho$.

Although the interval $\mathcal{I}_{\rho}$ does not possess an exact U(1) isometry, the endpoint $\rho^{\ast}$ is an (large) integer multiple of $2\pi$. Namely, it is topologically, but not metrically, equivalent to an $\text{S}^1$. In order to ensure the single-valuedness of the wave-function, we need to carefully impose boundary conditions. We choose $\Psi \big|_{\rho=0}=\Psi \big|_{\rho=\rho^{\ast}}$ and $\Psi' \big|_{\rho=0}=\Psi' \big|_{\rho=\rho^{\ast}}$. Note that this choice is, also, motivated by previous studies on related setups \cite{Speziali:2019uzn,Rigatos:2022ktp,Zhang:2024pim}. 
By focusing on the Hilbert space of such functions only, we can clearly see that the term $\Psi \Psi'\big|_0^{\rho^\ast}$ vanishes\footnote{We emphasize that there is another set of conditions that causes $\Psi \Psi'\big|_0^{\rho^\ast}$ to vanish at the endpoints, which is when $\Psi(0)=\Psi(\rho^{\ast})=0$. However, this choice is not consistent with \cref{eq: mode_minimal_sltn_final} and it also leads to the problem of making all the terms in \cref{integration equation} vanish. The concrete evidence that this term should vanish comes from a direct comparison with the solutions of \cite{Rigatos:2022ktp}; please see the comments below \cref{eq: mode_minimal_sltn_final}.}. We take into the fact that $h_4$, $h_8$, $u$, $n$, and $p$, are non-negative and we can deduce in a straightforward manner that:
\begin{equation}\label{unitarity bound}
        M^2\geq 4\ell(\ell+1).
\end{equation}    

The next step involves obtaining the ``minimal universal class of solutions''. The interpretation of ``universal'' in this context has already been elucidated. The term ``minimal'' is related to the fact that these solution saturate the bound obtained above for $M^2$, see \cref{unitarity bound}. We set $M^2 = 4\ell(\ell+1)$ in the Sturm-Liouville problem described by \cref{sturm_liouville standard form,eq: sturm_liouville} and it is reduced to:

%pertains to our objective of attaining the previously derived bound for $M^2$, as per \cref{unitarity bound}. Upon the establishment of $M^2 = \ell(\ell+1)$, the Sturm-Liouville problem delineated by \cref{sturm_liouville standard form,eq: sturm_liouville} is reduced to:
    \begin{equation}\label{eq: mode_minimal}
    \frac{\text{d}^2}{\text{d}\rho^2}\Psi=0\,          .
    \end{equation}
We can readily obtain the solution to the above differential equation. It is given by:
%that possesses a readily apparent solution
    \begin{equation}\label{eq: mode_minimal_sltn}
    \Psi=a\rho+\Psi_0\,          ,
    \end{equation}
In the expression above, $a$ and $\Psi_0$ are constants of integration. It is necessary to evaluate the general solution's conformity with the previously derived boundary conditions. A straightforward analysis demonstrates:
    \begin{equation}\label{eq: mode_minimal_sltn_final}
    \Psi = \Psi_0\,          .
    \end{equation}

The ``minimal universal class of solutions'' that characterizes these spin-2 fluctuations is simply given by a constant $\Psi_0$. It is important to note that this stands in stark contrast to the solutions for a linear function $u(\rho)$ that were derived in \cite{Speziali:2019uzn}. Truly, in \cite{Speziali:2019uzn}, the author used the same ansatz as the one we employed in this work, see \cref{eq: mode_expansion}, and examined this class of $\text{AdS}_3$ supergravity solutions for a linear $u(\rho)$. What is important to be stressed is that between our analysis here and the one performed in \cite{Speziali:2019uzn} the only difference is the form of $u(\rho)$, which is a consequence of the BPS identities. We ended up deriving that $\Psi = \Psi_0$, while in \cite{Speziali:2019uzn}, it was computed that $\Psi = u^{\ell} \Psi_0$, and in both cases $\Psi_0$ is a non-zero constant. In the case of a quiver defined by a linear function $u(\rho)$, there is a non-trivial dependence on $u$ itself at the level of the mode solution. 

Before we proceed we would like to make some comments connecting to the previous results of \cite{Speziali:2019uzn}. One could argue that given the solution derived in \cite{Speziali:2019uzn}, we could have considered the limit $u(\rho)$ goes to a constant at the level of having obtained the mode solutions. However, this approach could potentially be a bit risky without a careful examination of all the intermediate steps, particularly taking into consideration that the two setups are mathematically different. An important example of their differences is that in \cite{Speziali:2019uzn}, the author ended up with a singular Sturm-Liouville problem, whereas we have obtained an ordinary one. 

%The significance of spin-2 operators in AdS$_3$ holography is often understood through the additional fluxes in the NS-NS and R-R sectors of the theory, along with non-trivial warp factors in the geometry. However, our research proposes that the BPS conditions of the background supergravity theory play a more crucial role. This assertion is based on a detailed analysis of the $u=\text{constant}$ background given by \cref{eq: metrics}, in which we have observed that the solution to the BPS equations for $u$ is the only difference from previous studies.

The supergravity mode solutions that are dual to the metric fluctuations we have considered in this work are dual to operators in the field theory with conformal dimensions $\Delta$. We can use the well-established AdS/CFT relation between the mass, $M^2$, of the supergravity states and the conformal dimension of the field theory operators
    \begin{equation}
    M^2=\Delta(\Delta-2)
    \,          ,
    \end{equation}
and based on the previously derived inequality, see \cref{unitarity bound}, we can deduce that the scaling weight of the operators has a lower bound given by:
    \begin{equation}
    \Delta\geq\Delta_{\text{min}}, \qquad\text{with}\qquad  \Delta_{\text{min}}=2(\ell+1).
    \end{equation}

%We recognize the potential to investigate non-minimal solutions, encompassing cases where either $M^2>\ell(\ell+1)$ or substantial excitations manifest along T$^4$ or $\mathcal{I}_\rho$. However, an in-depth analysis of these latter solutions necessitates specific selections for $h_4$ and $h_8$, which are not pursued in this study.    
\section{The holographic central charge}\label{sec: cc}
In this section, we wish to present an approach for the computation of the holographic central charge for the theories defined by \cref{eq: nsns_sector_metric,eq: nsns_sector_full,eq: rr_sector_full}. In order to compute the holographic central charge, we should compute the normalization of the $2$-point function of the
operators that are dual to the graviton fluctuations that we studied earlier. As we saw in \cref{sec: ub&us} the ``universal minimal solution'' with no excitations along any of the directions of the internal manifold, namely $\ell=m=n=0$, is the massless graviton state and hence is dual to the stress-energy tensor. In order to obtain the normalisation of the $2$-point function for the stress-energy tensor, we have to read off the normalisation from the effective action of the three dimensional graviton.

The starting point for our computations in this section involves writing the massive type IIA action in the Einstein. It has the following schematic form:
\begin{equation}\label{eq: typeiib_action}
    S_{\text{IIA}}=\frac{1}{2\kappa_{10}^2}\int\sqrt{-g}(R+\cdots)\text{d}^{10}x
\,          .
\end{equation}
We can expand the action above to quadratic order in the metric fluctuations, which results in:
    \begin{equation}\label{eq: action_centrlacharge}
    \delta^2S=\frac{1}{\kappa_{10}^2}\int h^{\mu\nu}\partial_M\sqrt{-g}g^{MN}\partial_Nh_{\mu\nu}\text{d}^{10}x+\ldots
    \,      ,
    \end{equation}
where in the above we have used ``$\ldots$'' to represent a boundary term. We expand the term in the action explicitly and drop the boundary term to obtain:
    \begin{equation}
    \delta^2S[h_{\mu\nu}]=\frac{1}{\kappa_{10}^2}\int\sqrt{-g_{\text{AdS}_3}}\sqrt{\hat{g}}h^{\mu\nu}\left(\Box_{\text{AdS}_3}^{(2)}+2+\hat{\Box}\right) h_{\mu\nu}\text{d}^{10}x
    \,          .
    \end{equation}
We note that the operator $\hat{\Box}$ is explicitly defined in equation \ref{eq: hatbox_def}. We can use the ansatz employed previously for the spin-2 modes, $h_{\mu \nu}$, that we re-state for the convenience of the reader
    \begin{equation}
    h_{\mu\nu}=\sum_{\ell mn}(h_{\ell mn}^{(tt)})_{\mu\nu}Y_{\ell m}\Psi_{\ell mn}e^{in\cdot\theta}
    \end{equation}
in the action and we obtain 
    \begin{equation}
    \delta^2S=\sum_{\ell mn}C_{\ell mn}\int_{\text{AdS}_3}\sqrt{-g_{\text{AdS}_3}}(h_{\ell mn}^{(tt)})^{\mu\nu}\left(\Box_{\text{AdS}_3}^{(2)}+2-4\ell(\ell+1)\right)(h_{\ell mn}^{(tt)})_{\mu\nu}\text{d}^2x
    \end{equation}
where the coefficients $C_{\ell mn}$ are:
    \begin{equation}\label{eq: clmn_def}
    C_{\ell mn} = \frac{16\pi^5}{\kappa_{10}^2} \int_{\mathcal{I}_{\rho}} \sqrt{\hat{g}} (e^{-\frac{\Phi}{2}} f_1)^{\frac{1}{2}} |\Psi_{\ell mn}|^2\text{d}\rho
    \,          ,
    \end{equation}
and we have used the standard normalisation for the spherical harmonics
    \begin{equation}
    \int Y_{\ell m}Y_{\ell'm'}\text{d}^2x=1
\,          .
    \end{equation}

\Cref{eq: clmn_def} is finite for the fluctuations we have considered in this work, as they remain finite at all points of the background spacetime. We focus on the ``minimal universal solution'' as derived in \cref{sec: ub&us}. By setting all quantum numbers to zero, we can assign $|\Psi_{000}|^2=1$, resulting in:
\begin{equation}\label{central charge}
C_{000}=\frac{16\pi^5}{\kappa_{10}^2}\int_0^{\rho^\ast}h_4h_8\text{d}\rho,
\end{equation}
where in deriving the above we have used the standard results:
    \begin{equation}
     \text{vol}_{\text{S}^2}=4\pi,
     \quad 
     \text{vol}_{\text{T}^4}=(2\pi)^4, 
     \quad
     \text{vol}_{\mathcal{I}_\psi}=2\pi,
     %\quad
     %2\kappa_{10}^2=(2\pi)^7.
    \end{equation}
The aforementioned outcome, as \cref{central charge}, aligns with the calculation of holographic entanglement entropy presented in the work by \cite{Lozano:2019zvg}, up to an unimportant numerical factor. We note that in spite of our class of ``universal minimal solution'' being different to the one derived in \cite{Speziali:2019uzn} the expressions for the holographic central charge agree exactly, see \cref{central charge} and \cite[equation (37)]{Speziali:2019uzn}. 

%It is important to note that using the term ``Entanglement Entropy'' may not be precise in a quantum mechanical system, and a more appropriate term would be ``the volume of the internal space''. Nonetheless, this is consistent with the entanglement entropy in the higher-dimensional system as discussed in \cite{Klebanov:2007ws,Macpherson:2014eza}, and we have adopted this term in our case as well, albeit with a slight deviation from the standard nomenclature.

Before closing this section we feel that it is necessary to make the following comment for completeness and clarity. In a general AdS supergravity solution, gravitons should be treated carefully as they can be distinct from massive spin-2 fields, see as a representative example \cite{PhysRevD.32.389} where the authors explain and solve the mixing problem. However, it is important to note that the normalization of the $3$-dimensional massless graviton can still be obtained, without issues, by taking the $\ell=m=n=0$ solution. 

\section{Two-dimensional superconformal multiplets}\label{sec: multiplets}

Using standard techniques from the work of \cite{Cordova:2016emh,Cordova:2016xhm} related to superconformal field theories in $D \geq 3$ dimensions, the authors of \cite{Kos:2018glc} worked out the spectrum of superconformal multiplets in two-dimensional theories that realize $\mathcal{N}=(4,0)$ supersymmetry. This is important for our purposes as it allows us to identify the field-theory operators dual to the spin-2 fluctuations in the bulk that we computed earlier. We stress again that in this work we are interested in theories for which $u = \text{constant}$ holds. The analysis of this section for the case of $u = \text{constant} \rho$ has been performed in \cite{Speziali:2019uzn}. As we will see the final result coincides with that of \cite{Speziali:2019uzn}. This is just the statement that while the supergravity modes are different in those setups, the field theory operators that are mapped to those modes have the same form.

For each one of the metric fluctuations computed in the previous sections there exists a gauge invariant operator in the boundary field theory. The states of the two-dimensional $\mathcal{N}=(4,0)$ superconformal algebra can be labelled by the scaling weights of the $\text{SL}(2,\mathbb{R}) \times \overline{\text{SL}(2,\mathbb{R})}$ conformal algebra which we denote by $h, \overline{h}$ respectively. There is, also, the label, $R$, of the $\text{SU}(2)_{\text{R}}$ R-symmetry. The conformal dimension of an operator is given as the sum of the scaling weights, $\Delta = h + \overline{h}$, and its spin is taken to be their difference $s = h - \overline{h}$. We, therefore, schematically represent a state of the superconformal algebra in the following way:
\begin{equation}\label{eq: schematic_operator}
[h,\overline{h}]^{R}_{\Delta} 
\,          .
\end{equation}

In the bulk supergravity description, the R-symmetry is realized as the isometries of the two-dimensional sphere and hence, the quantum number associated with it, $\ell$, is related to the charge of the R-symmetry of the field theory operators; in particular $R = 2 \ell$, such that the R-symmetry charge is always a non-negative, even integer. 

We have already established that for the minimal universal solution the corresponding conformal dimension is given by $\Delta = 2 (\ell+1)$. Finally, we stress that the scaling weights of the conformal algebra are not independent for the type of fluctuations that we are considering in this work, but rather they obey the relation $h - \overline{h}=\pm 2$, see e.g \cite{Speziali:2019uzn}.

After taking all the above into consideration, the schematic form presented in \cref{eq: schematic_operator} becomes 
\begin{equation}
[h,h\pm 2]^{2\ell}_{2h\pm 2} 
\,          .
\end{equation}
We observe that the choice $h=0$ leads to $[0,2]^{0}_{2}$. These are the quantum numbers of the anti-holomorphic stress-energy tensor. The choice $h=2$ leads to $[2,0]^{0}_{2}$, which are the quantum numbers of the holomorphic stress-energy tensor. As computed in \cite{Kos:2018glc} this state arises as a top component descendant in a short multiplet with a conformal primary that has the quantum nmubers $R=2$ and $h=1$. 

Before closing off this section we wish to make a brief comment about massive fluctuations resulting from our spin-2 analysis. These are solutions for which $\ell \neq 0$. These supergravity modes correspond to operators that are top components in a superconformal multiplet with a primary operator of dimension $\Delta = 2 \ell +1$ or to operators obtained as combinations of chiral primaries in short multiplets with operators in the anti-holomorphic part of the algebra, see \cite{Kos:2018glc} and also \cite{Speziali:2019uzn} for a more thorough discussion on this.

\section{Discussion and outlook}\label{sec: final}
In this work we examined spin-2 fluctuations around a warped solution within the Type IIA theory that is of the schematic form $\text{AdS}_3 \times \text{S}^2 \times \text{T}^4 \times \mathcal{I}_\rho$. This solution involves fluxes in the NS-NS and R-R sectors. One motivation was to check explicitly if the different choice for the function $u(\rho)$, namely either it being a simple constant or a linear function, would lead to a drastically different class of solutions as it happened to the related $\text{AdS}_2$ backgrounds \cite{Rigatos:2022ktp,Zhang:2024pim}. Another was to fully understand and establish the map presented in \cref{fig：holographic central charge}, relating the spin-2 metric fluctuations and expressions for the holographic central charges in backgrounds that either related by a different choice of the $u(\rho)$ function or via a T-duality transformation, see \cite{Lozano:2019jza,Lozano:2020txg} for detailed discussion on the precise relation of these backgrounds. As we have discussed in the previous sections, the expressions for the holographic central charge are the same in all those theories and it turns out that one can pass from the spin-2 mode solutions of the case $u(\rho)$ being a linear function to the case $u(\rho)$ is a constant without any problems, while the reverse does not seem to be possible. 

%Following the approach outlined in a previous study (\cite{}), we derived the equations of motion that govern the behavior of these spin-2 states. Consistent with previous findings, we categorized these states into two main classes: universal and non-universal. It is noteworthy that the universal solutions are not influenced by the defining functions of the background and are therefore present in all 2-dimensional field theories that are dual to the backgrounds studied here. 

In our work, we specifically computed the ``minimal universal class of solutions'' which represents a set of solutions that saturates the unitarity bound and does not depend on the specific definitions of the characteristic functions that determine the background. These supergravity states act as sources for operators in the dual field theory, with a classical dimension of $\Delta=\ell+2$, where $\ell$ represents the quantum number on the $\text{S}^2$. This is the geometrical realization of the $\text{SU}(2)_R$ R-symmetry.

By utilizing this class of solutions and the supergravity action expansion, we were able to determine the expression for the holographic central charge. We verified that it aligns with the calculation of the holographic entanglement entropy undertaken in a previous study \cite{Lozano:2019zvg}, as expected. This provides us with a good consistency check.

%We would like to highlight that our work complements the findings of \cite{Speziali:2019uzn}. The main distinction between these two works lies in the choice of the function $u(\rho)$. Specifically, we have chosen it to be a constant, while \cite{Speziali:2019uzn} considered a linear function. This difference in the two choices of $u(\rho)$ manifests in the solutions of the metric perturbations $h_{\mu\nu}(x,z)$. More precisely, we have determined that $u(\rho)=\text{constant}$ results in a constant solution, whereas a non-trivial functional dependence of $h_{\mu\nu}(x,z)$ is only achieved when $u(\rho)$ is a linear function. It is worth noting that the expressions for the unitarity bound and central charge are the same in both cases.

%Our findings lead to the following interpretation. In backgrounds with pure $\text{AdS}_3$ factors, it was previously established that the modes under consideration are non-trivial only when the mass is zero, as discussed in e.g., \cite{}. Recent explorations of $\text{AdS}_3$ supergravity solutions indicate that non-trivial warp factors and fluxes contribute to a more intricate and exotic description, as suggested in \cite{}. However, our choice of $u$ resulted in trivial spin-2 states without requiring the warp factors and fluxes to be constant. Therefore, we conclude that the supersymmetry conditions play a more crucial role.

Before concluding this section, we would like to highlight some intriguing avenues for future research work.

Firstly, examining the most general solutions for which the function $h_4$ admits support both on the $\rho$-dimension as well as the $\text{T}^4$ is an interesting avenue of future work. While it is conceptually straightforward, this would lead necessarily to a more complicated equation, namely a partial differential equation, describing the dynamics of the corresponding spin-2 states. 
    
\begin{figure}[H]
\subfloat[A background with a constant function $u(\rho)$. The functions $h_4$ and $h_8$ both vanish at the beginning of the $\rho$-dimension, however, $h_8$ is non-vanishing at the other endpoint.]{\includegraphics[width = 0.45\textwidth]{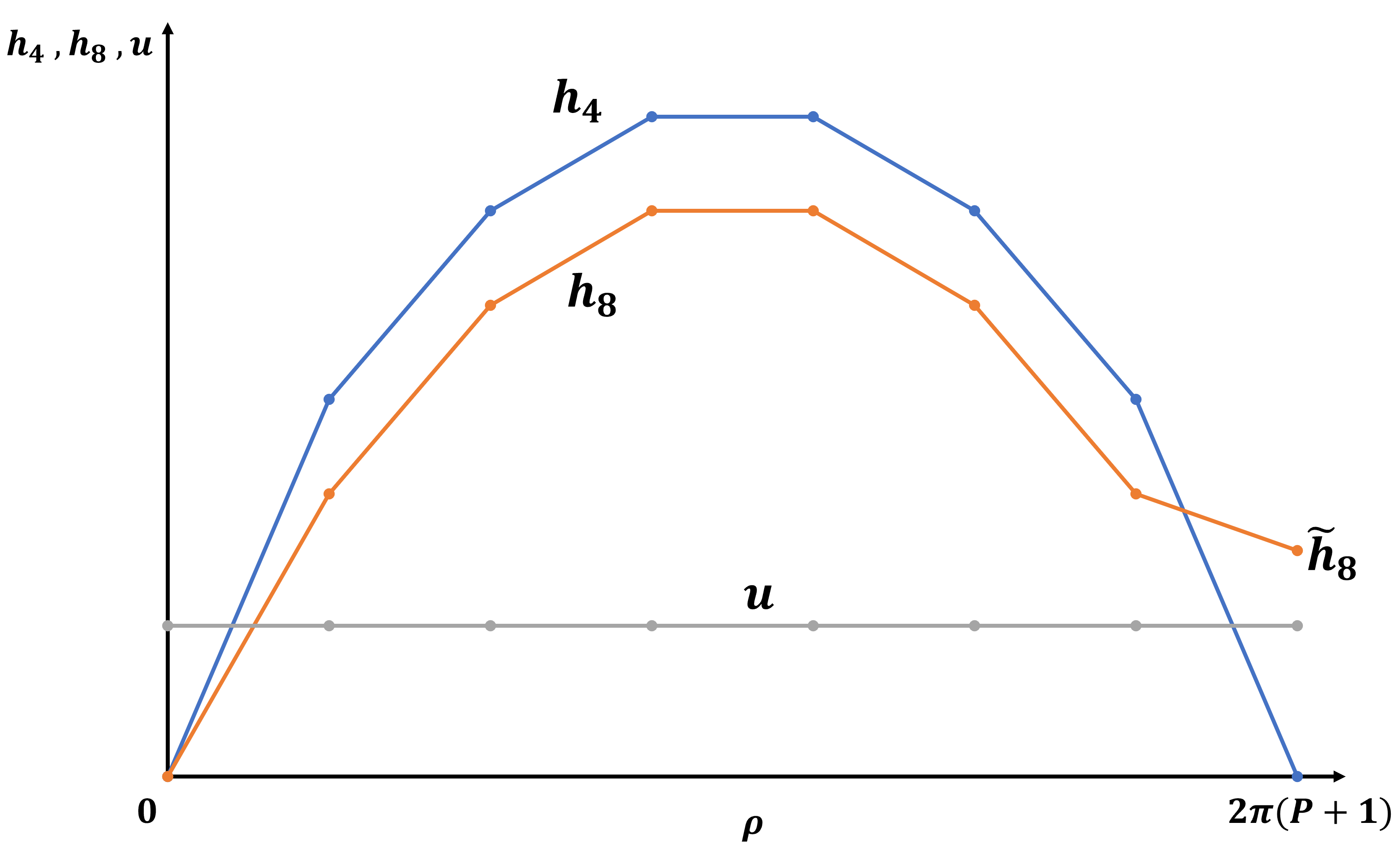}\label{fig: a}} \hfill
\subfloat[A background with a constant function $u(\rho)$. The functions $h_4$ and $h_8$ both vanish at the final endpoint of the $\rho$-coordinate, however, $h_4$ does not vanish at the beginning.]{\includegraphics[width = 0.45\textwidth]{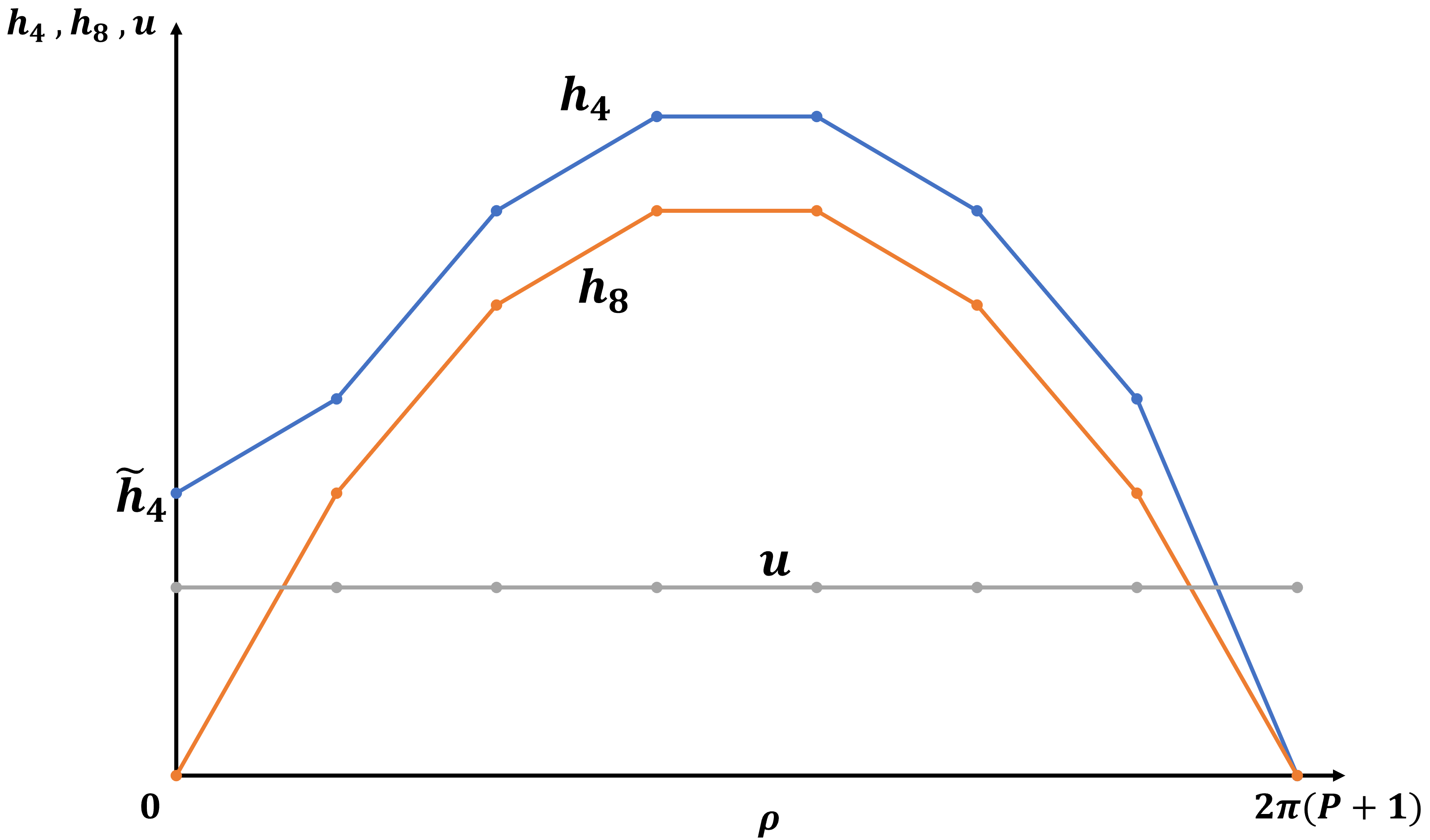}\label{fig: b}}\\
\subfloat[A background with a linear function $u(\rho)$. The functions $h_4$ and $h_8$ both vanish at the beginning of the $\rho$-dimension, however, $h_8$ is non-vanishing at the endpoint.]{\includegraphics[width = 0.45\textwidth]{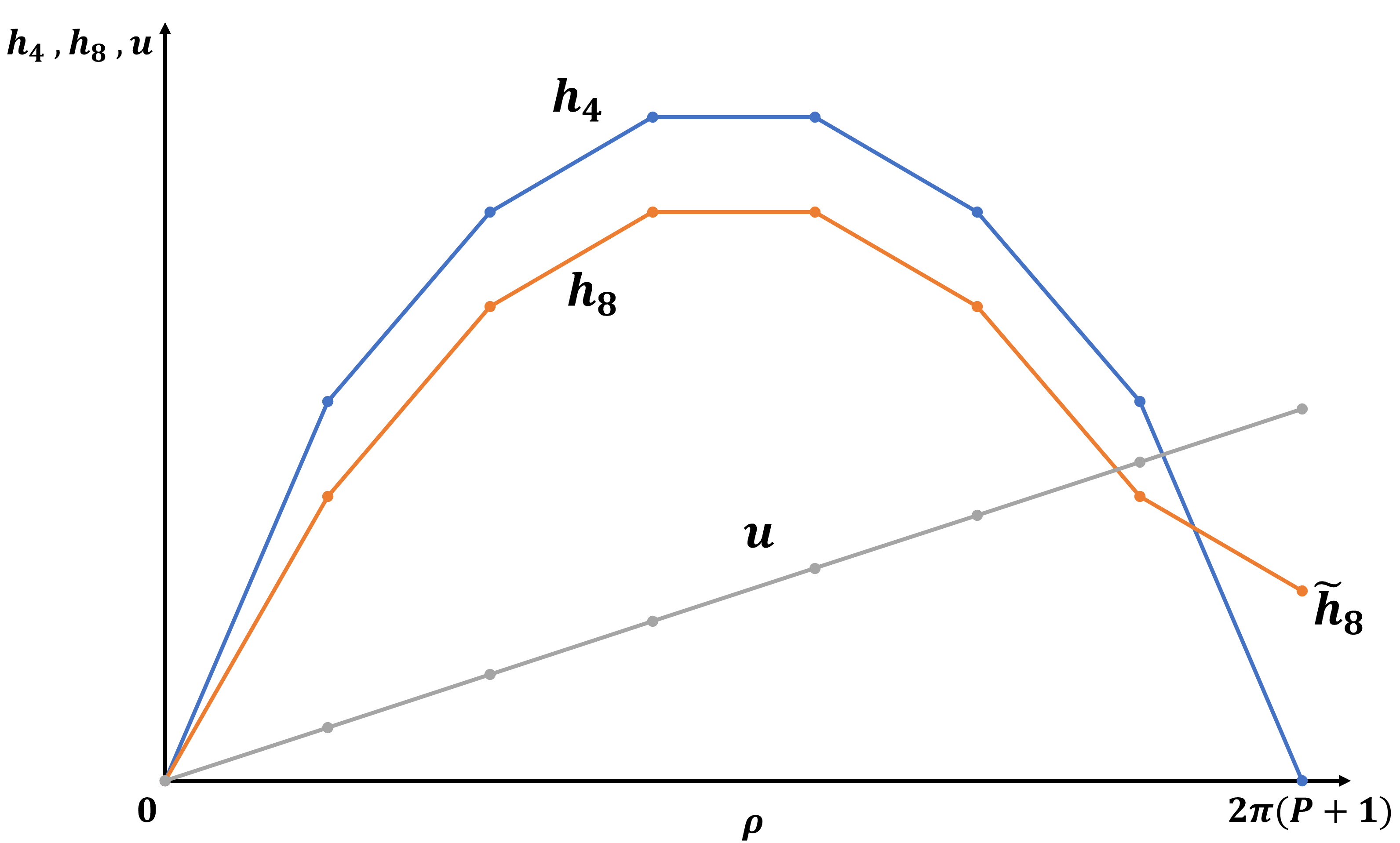}\label{fig: c}} \hfill
\subfloat[A background with a linear function $u(\rho)$. The functions $h_4$ and $h_8$ both vanish at the beginning of the $\rho$-coordinate, however, $h_4$ does not vanish at the endpoint.]{\includegraphics[width = 0.45\textwidth]{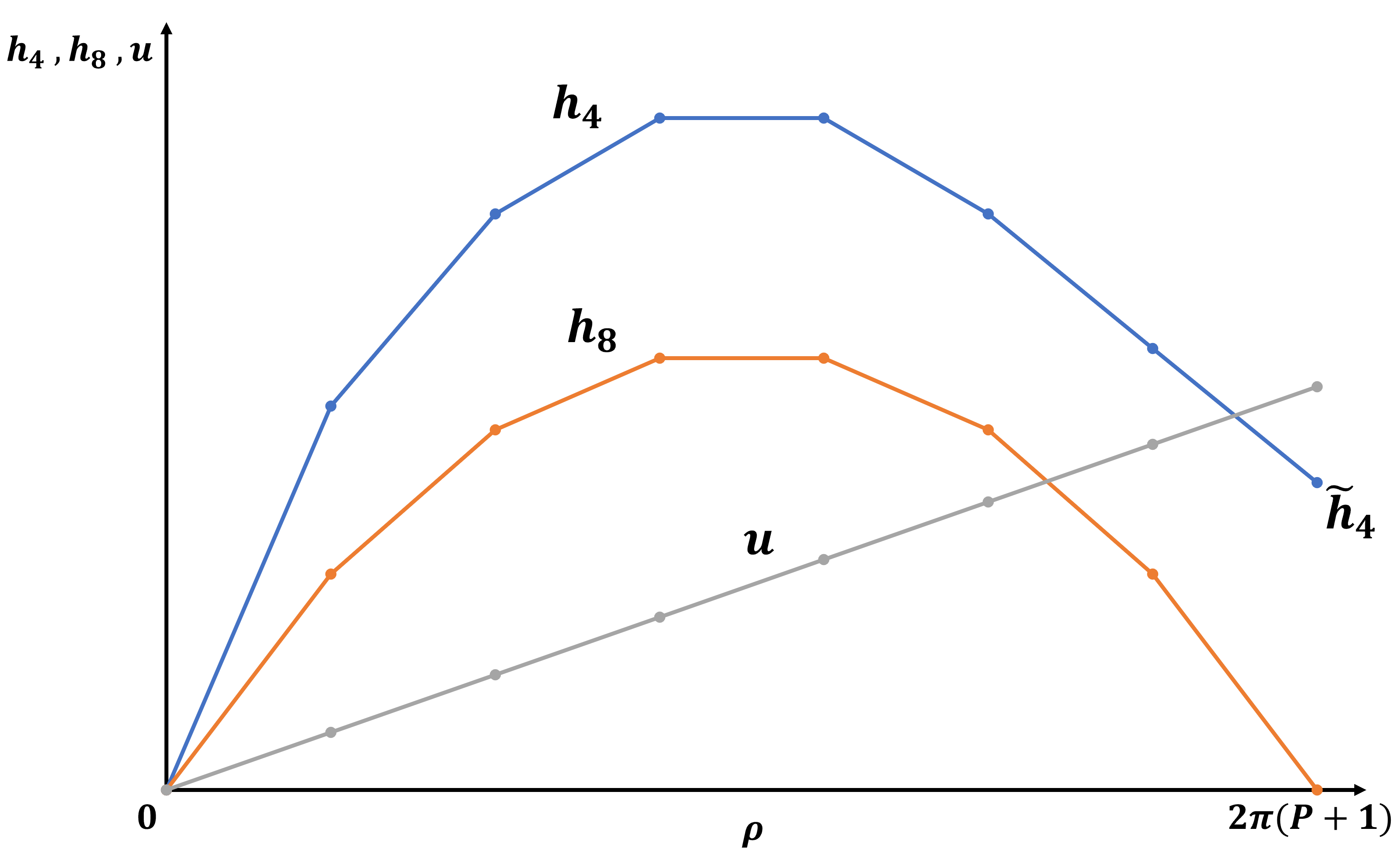}\label{fig: d}} 
\caption{Examples of supergravity backgrounds defined by piecewise linear $h_4$ and $h_8$ functions. In \cref{fig: a,fig: b} the roles of $h_4$ and $h_8$ can be freely interchanged. In \cref{fig: c,fig: d} where $u(\rho)$ is a linear function, both $h_4$ and $h_8$ have to vanish at the beginning of ${\rho}$-dimension. These conditions on the behavior of the defining $h_4$ and $h_8$ functions are such that the internal space closes properly on the $\rho$-dimension \cite{Filippas:2020qku}.}
\label{quivers2}
\end{figure}

Furthermore, in this work we considered the backgrounds derived in \cite{Lozano:2019jza} with the special condition that both $h_4$ and $h_8$ vanish at the endpoints of the interval $\mathcal{I}_{\rho}$. We further imposed that $u(\rho) = \texttt{constant}$, while \cite{Speziali:2019uzn} considered the case $u(\rho) = \tfrac{b_0}{2\pi} \rho$, with $b_0$ being a constant. Hence, we have obtained the spin-2 part of the KK spectra for the quivers described pictorially in \cref{quivers1}:
\begin{figure}[H]
\subfloat[A toy example of piecewise linear functions for the definitions of $h_4, h_8$ and $u$ that define a specific supergravity background. In the example depicted here both $h_4$ and $h_8$ vanish at the endpoints of $\mathcal{I}_{\rho}$.\label{fig: sub_1}]{\includegraphics[width = 0.45\textwidth]{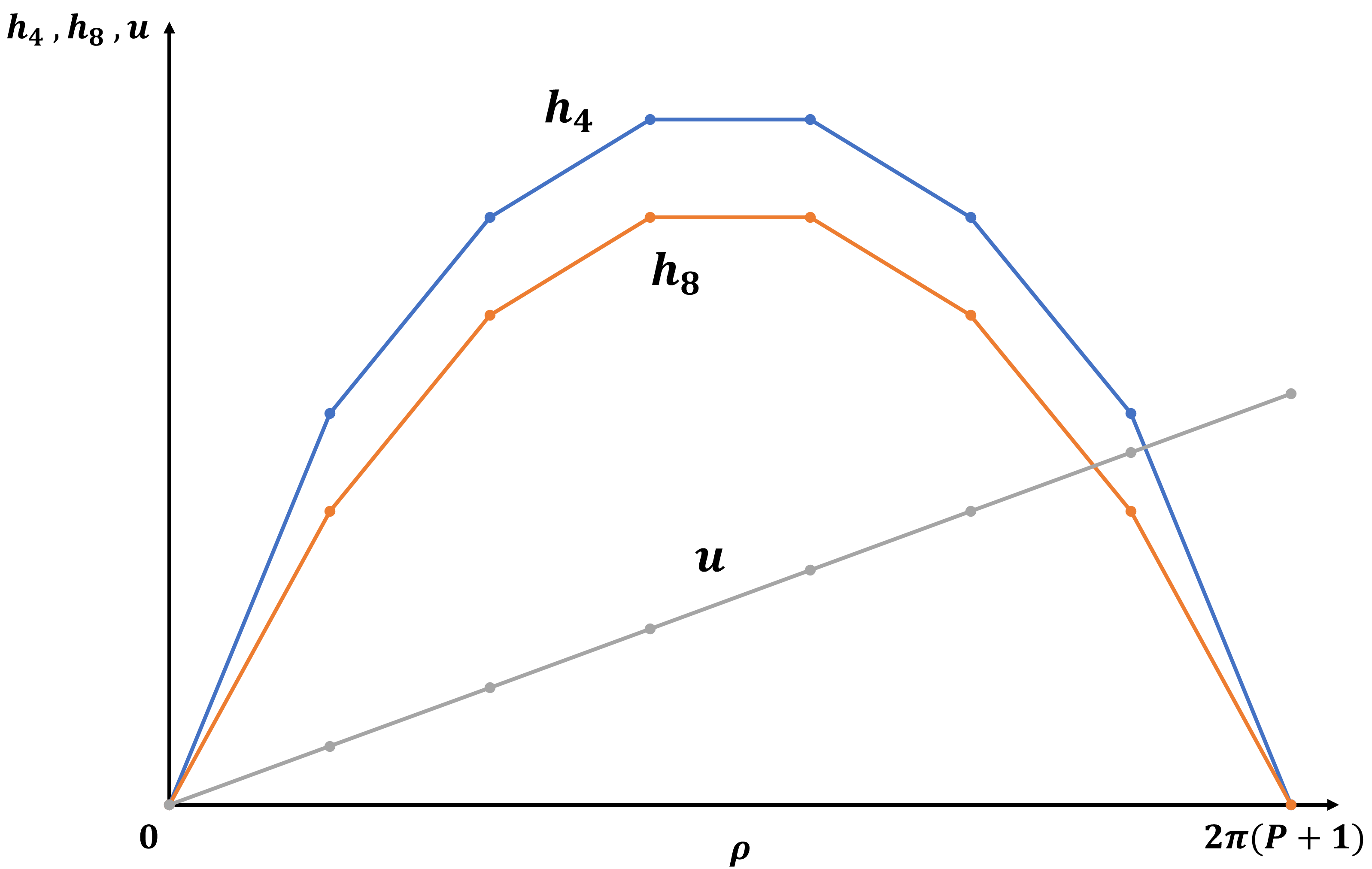}} \hfill
\subfloat[An example of a supergravity background given by piecewise linear $h_4$ and $h_8$ functions, while $u$ is defined to be a context. In this example both $h_4$ and $h_8$ vanish at the endpoints of $\mathcal{I}_{\rho}$.\label{fig: sub_2}]{\includegraphics[width = 0.45\textwidth]{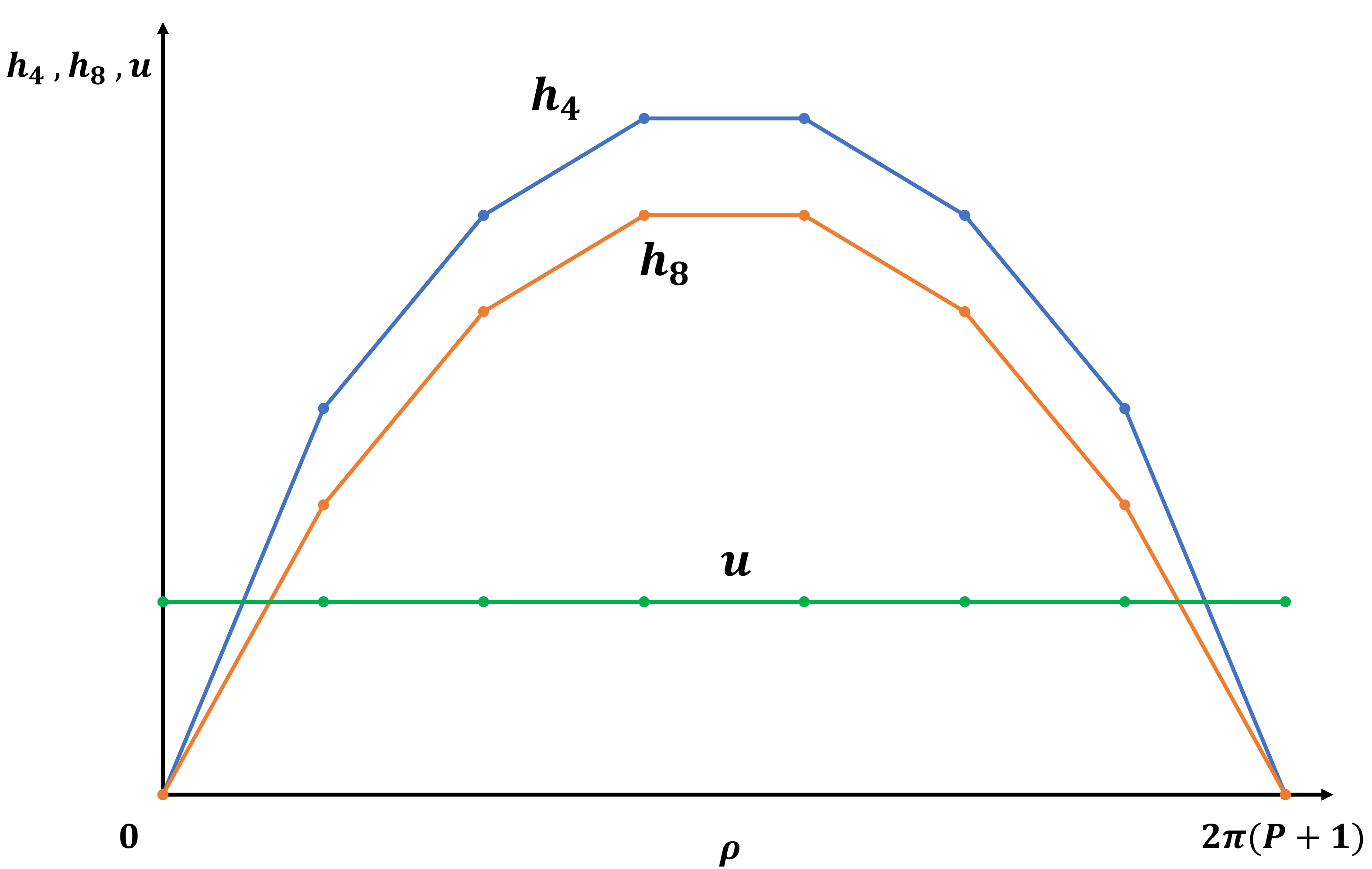}}
\caption{Examples of supergravity backgrounds defined by piecewise linear $h_4$ and $h_8$ functions such that they vanish at the endpoints of the $\rho$-interval. In \cref{fig: sub_1} we have a linear $u(\rho)$ function, while in \cref{fig: sub_2} $u(\rho)$ is given by a constant.}
\label{quivers1}
\end{figure}
It would be interesting, and challenging, to examine quivers for which the functions $h_4$ and $h_8$ do not vanish at the beginning and end of the $\rho$-interval, see \cref{quivers2} for a schematic pictorial depiction. The immediate and practical complication that we can already spot has to do with picking boundary conditions for the wave-functions appropriately in these cases. Other than that, the rest of the analysis we have presented in this work should applicable in a straightforward manner.
The kind of quivers depicted via their supergravity backgrounds in \cref{quivers2} have been thoroughly analysed in \cite{Filippas:2020qku}. They were found to be consistent with anomaly cancellation from the field theory point of view and the dual supergravity backgrounds are consistent with the BPS equations and Bianchi identities of massive type IIA. 

%Finally, another related approach would be to, also, consider metric perturbations along the $\text{T}^4$ directions of the supergravity background. This is an interesting topic as it would provide us with further information on the KK-spectrum of those theories. Note, of course, that while the computational part of such an endeavour would be similar to the above, the field theory interpretation is completely different as such modes correspond to spin-0 operators in the dual field theory.
%%%%%%%%%%%%%%%%%%%%%%%%%%%%%%%%%%%%%%%%%%%%%%%%%%%%%%%%%%%%%%%%%%%%%%%%%%%%%%%%%%%%%%%%%%%%%%%%%%%%%%%%%%
\newpage
%%%%%%%%%%%%%%%%%%%%%%%%%%%%%%%%%%%%%%%%%%%%%%%%%%%%%%%%%%%%%%%%%%%%%%%%%%%%%%%%%%%%%%%%%%%%%%%%%%%%%%%%%%
\section*{Acknowledgments} 
We thank Kostas Rigatos and Xinan Zhou for feedback on the draft and for help with the English language in writing this paper. The work of Shuo Zhang is supported by funds from the Kavli Institute for Theoretical Sciences (KITS).
%%%%%%%%%%%%%%%%%%%%%%%%%%%%%%%%%%%%%%%%%%%%%%%%%%%%%%%%%%%%%%%%%%%%%%%%%%%%%%%%%%%%%%%%%%%%%%%%%%%%%%%%%%
\newpage
%%%%%%%%%%%%%%%%%%%%%%%%%%%%%%%%%%%%%%%%%%%%%%%%%%%%%%%%%%%%%%%%%%%%%%%%%%%%%%%%%%%%%%%%%%%%%%%%%%%%%%%%%%
\bibliographystyle{ssg}
\bibliography{spin2bib}
%%%%%%%%%%%%%%%%%%%%%%%%%%%%%%%%%%%%%%%%%%%%%%%%%%%%%%%%%%%%%%%%%%%%%%%%%%%%%%%%%%%%%%%%%%%%%%%%%%%%%%%%%%
%%%%%%%%%%%%%%%%%%%%%%%%%%%%%%%%%%%%%%%%%%%%%%%%%%%%%%%%%%%%%%%%%%%%%%%%%%%%%%%%%%%%%%%%%%%%%%%%%%%%%%%%%%
\end{document}